\documentclass[reprint,aps,pre,showpacs,showkeys,longbibliography]{revtex4-2}
\usepackage{amsmath,amsfonts,graphicx,amssymb,array,natbib}
\usepackage[caption=false]{subfig}
\usepackage{hyperref,bm}
\usepackage{csquotes}
\begin{document}

\title{Exact transparent boundary condition for the multidimensional Schr\"odinger equation in hyperrectangular computational domain}

\author{R.M. Feshchenko},
\affiliation{P.N. Lebedev Physical Institute of RAS, 53 Leninski Pr., Moscow, Russia, 119991}
\email{rusl@sci.lebedev.ru}
\author{A.V. Popov}
\affiliation{Pushkov Institute of Terrestrial Magnetism, Ionosphere and Radiowave Propagation of RAS, Troitsk, Moscow region, Russia, 142190}

\date{\today}

\begin{abstract}
In this paper an exact transparent boundary condition (TBC) for the multidimensional Schr\"odinger equation in a hyperrectangular computational domain is proposed. It is derived as a generalization of exact transparent boundary conditions for 2D and 3D equations reported before. A new exact fully discrete (i.e. derived directly from the finite-difference scheme used) 1D transparent boundary condition is also proposed. Several numerical experiments using an improved unconditionally stable numerical implementation in the 3D space demonstrate propagation of Gaussian wave packets in free space and penetration of a particle through a 3D spherically asymmetrical barrier. The application of the multidimensional transparent boundary condition to the dynamics of the 2D system of two non-interacting particles is considered. The proposed boundary condition is simple, robust and can be useful in the field of computational quantum mechanics, when an exact solution of the multidimensional Schr\"odinger equation (including multi-particle problems) is required.
\end{abstract}

\maketitle

\section{Introduction} 

The methods used to solve the multidimensional time dependent Schr\"odinger equation (TDSE) in finite computational domains require suitable boundary conditions that truncate the wave function at the domain's boundaries. Among many types of the known boundary conditions the exact transparent boundary conditions (TBC) stand out because they provide the most precise solution to this boundary truncation problem while potentially preserving the stability of the finite-difference scheme used \cite{ComCompPhys_Antoine_2008}. There exist many different forms of TBCs, although the majority of them are one-dimensional, for instance, a widely known Baskakov--Popov TBC \cite{popov1996accurate}. A number of 2D and 3D TBCs \cite{ComCompPhys_Antoine_2008,PhysRev_Heinen_2009} have also been known for some time but not for (hyper)rectangular computational domains, which are the most convenient in practice.

In our previous papers we reported a number of exact transparent boundary conditions. Among them are: a TBC for the 3D parabolic wave equation \cite{feshchenko2011exact} in the rectangular domain, which is equivalent to the 2D TDSE, and a TBC for the 3D Schr\"odinger equation \cite{feshchenko2013exact} in the rectangular cuboid domain. In this paper we generalize these conditions to the Schr\"odinger equation with arbitrary number of spatial dimensions and in the potential of quite a general form. 

Using thus derived multidimensional TBC we then proceed to the solution of a number of model problems in the 3D space by applying a new finite-difference (FD) implementation of the 3D exact transparent boundary condition based on the exact unconditionally stable fully discrete 1D TBC and the iterative inversion of sparse matrices. Both the 1D fully discrete TBC and efficient sparse matrix inversion method allow us to significantly reduce the computational costs while simultaneously increasing the computational precision. The specific numerical experiments with a sum of Gaussian wave packets and of a particle penetrating through a 3D spherically asymmetrical barrier are carried out and discussed. In addition to the 3D case, we briefly consider the numerical implementation of the proposed TBC in the 4D space corresponding to the two-particle Schr\"odinger equation in two dimensions.

\section{Multidimensional transparent boundary condition} 
\label{secmult}
\subsection{Multidimensional Schr\"odinger equation}
In order to derive a general TBC in the N-dimensional hyperrectangular computational domain let us consider a multidimensional time-dependent Schr\"odinger equation in the following form
\begin{equation}
i\frac{\partial \psi}{\partial t}=-\Delta\psi+V(t,x_1\ldots x_N)\psi,
\label{2a}
\end{equation}
where potential $V$ can be expressed as a sum of single particle potentials
\begin{equation}
V(x_1\ldots x_N)=\sum\limits_{i=1}^{M}V_i(t,x_{d(i-1)+1}\ldots x_{di}).
\label{2b}
\end{equation}
Here $N=dM$, $M$ is the number of particles, $d$ is the physical space dimension, and $N$ is the total dimension of the configuration space $R^N$. Equation (\ref{2a}) describes the evolution of a system of $M$ different non-interacting particles in an external potential, which is not necessary vanishing at the infinity. It is relevant for various problems of quantum mechanics including those related to the evolution of the entangled states. The potential in the form \eqref{2b} can also be used to describe evolution of a single-particle system when the potential is a sum of potentials $V_i$ depending on subsets of the coordinates. 

As the next step, we will consider a single particle auxiliary equation in a d-dimensional space
\begin{equation}
i\frac{\partial \varphi}{\partial t}=-\Delta\varphi+U(t,x_1\ldots x_d)\varphi,
\label{2c}
\end{equation}
where $U$ is some potential. It is known \cite{brandt1991free} that any solution of ({\ref{2c}) can be expressed in terms of its propagator $\Gamma(t,t',x_1\ldots x_d, x'_1\ldots x'_d)$ and the initial value of the wave function at some moment $t=t'$ as
\begin{multline}
\varphi(t,x_1\ldots x_d)=\\
\int\limits_{-\infty}^{+\infty}\ldots\int\limits_{-\infty}^{+\infty}\Gamma(t,t',x_1\ldots x_d, x'_1\ldots x'_d)\times\\
\varphi(t',x'_1\ldots x'_d)dx'_1\ldots dx'_d.
\label{2d}
\end{multline}
Propagator $\Gamma$ considered to be a function of $t$ and $x_1\ldots x_d$ must satisfy the equation (\ref{2c}) and the following initial condition at $t=t'$
\begin{equation}
\Gamma(t,t,x_1\ldots x_d, x'_1\ldots x'_d)=\prod\limits_{i=1}^{d}\delta(x_i-x'_i).
\label{2e}
\end{equation}
The propagator will depend only on the difference $t-t'$ when potential $V$ is time-independent. Differentiating \eqref{2d} with respect to $t'$ and integrating by parts it can be shown that the propagator also satisfies the following \enquote{conjugate} equation
\begin{equation}
i\frac{\partial \Gamma}{\partial t'}=\Delta'\Gamma-U(x'_1 \ldots x'_d)\Gamma,
\label{2g}
\end{equation}
where Laplacian $\Delta'$ implies differentiation by $x'_1\ldots x'_d$. The equation \eqref{2g} will, in fact, coincide with \eqref{2c} if the propagator $\Gamma$ depends only on the difference $t-t'$.

Let us now introduce new function $\Phi$ defined as
\begin{multline}
\Phi(t_1,\ldots,t_M,x_1,\ldots,x_N)=\\
\int\limits_{-\infty}^{+\infty}\ldots\int\limits_{-\infty}^{+\infty}\psi(t_M,\zeta_1\ldots\zeta_{N-d},x_{N-d+1}\ldots x_N)\times\\
\prod\limits_{i=1}^{M-1}\Gamma_i(t_i,t_M,x_{(i-1)d+1}\ldots x_{id},\zeta_{(i-1)d+1}\ldots \zeta_{id})\times\\
d\zeta_{(i-1)d+1}\ldots d\zeta_{id},
\label{2h}
\end{multline}
where each propagator $\Gamma_i$ satisfies equation (\ref{2c}) with $U$ replaced by the potential $V_i$. The function $\Phi$ depends on $M$ variables $t_i$ that will be called here time-like variables, in addition to $N$ spatial coordinates. 

It can be demonstrated that $\Phi$ satisfies $M$ single-particle differential equations, instead of one equation (\ref{2a}). These equations are
\begin{multline}
i\frac{\partial \Phi}{\partial t_i}=\\
\sum\limits_{j=1}^{d}\left[-\frac{\partial^2\Phi}{\partial x_{(i-1)d+j}^2}+V_i(t_i,x_{(i-1)d+1}\ldots x_{id})\Phi\right],\label{2i}
\end{multline}
where $i=1\ldots M$. The first $M-1$ equations directly follow from the definition of $\Gamma_i$. The last equation (when $i=M$) can be obtained by differentiating (\ref{2h}) by $t_M$ then using (\ref{2c}), (\ref{2g}) and integrating by parts. Function $\Phi$ is related to the wave function $\psi$ by the following equality
\begin{equation}
\Phi(t_1=t,\ldots t_M=t,x_1,\ldots x_n)=\psi(t,x_1,\ldots x_n).
\label{2j}
\end{equation}
In every equation of (\ref{2i}) there is only one time-like variable and $d$ spatial coordinates. All other temporal and spatial variables are merely parameters. This makes it possible to simplify the transparent boundary condition for the main equation (\ref{2a}).

\subsection{Transparent boundary condition}
Let us assume that we are presented with a problem of solving equation (\ref{2a}) in a hyper-rectangular computational domain $\Xi$ defined as 
\begin{equation}
|x_i|<a_i,\quad i=1\ldots N,
\label{3a}
\end{equation}
where $a_i$ are positive numbers. The wave function at $t=0$ is supposed to be zero outside this computational domain and at its boundary $\partial\Xi$. The boundary condition should be linear with respect to the wave function and should relate the projection of its gradient on the normal of $\partial\Xi$, at time $t$, to its values on the boundary at all moments preceding $t$. Thus, in the most general form, a TBC for the equation (\ref{2a}), taken at a hyperfacet at $x_i=\pm a_i$, can be represented as
\begin{widetext}
\begin{equation}
\frac{\partial\psi}{\partial x_i}=\int\limits_{0}^{t}\int\limits_{\partial\Xi}K_i(t,\xi,x_1\ldots x_{i-1},\pm a_i,x_{i+1}\ldots x_N, \sigma_1\ldots \sigma_{i-1},\sigma_{i+1}\ldots \sigma_N)\psi(\xi,\sigma_1\ldots \sigma_{i-1},a_i,\sigma_{i+1}\ldots \sigma_N)d\Sigma_j\,d\xi,
\label{3b}
\end{equation}
\end{widetext}
where the spatial integral ($d\Sigma$) by variables $\sigma_{j\ne i}$ is taken over the boundary $\partial\Xi$ of a hyper-rectangular domain, and functions $K_i$ are some kernels that are not known \textit{a priory} and may be, in fact, difficult to obtain analytically as they depend on the complete potential $V$.

Similar to $\psi$, one can derive the following TBC for the function $\Phi$ at the i-th hyperfacet, using j-th equation of (\ref{2i})
\begin{widetext}
\begin{multline}
\frac{\partial\Phi}{\partial x_i}=\int\limits_{0}^{t}\int\limits_{\partial\Xi_j}\tilde K_i(t_j,\xi,x_{d(j-1)+1}\ldots x_{i-1},a_i,x_{i+1}\ldots x_{dj}, \sigma_{d(j-1)+1}\ldots \sigma_{i-1},\sigma_{i+1}\ldots \sigma_{dj})\times\\
\Phi(t_1\ldots t_{j-1},\xi,t_{j+1}\ldots t_M,x_1\ldots x_{d(j-1)},\sigma_{d(j-1)+1}\ldots \sigma_{i-1},a_i,\sigma_{i+1}\ldots \sigma_{dj},x_{dj+1}\ldots x_N)d\Sigma\,d\xi.
\label{3c}
\end{multline}
\end{widetext}
Here, $d(j-1)+1\le i\le dj$, and the spatial integral ($d\Sigma_j$) is taken only over the intersection $\partial\Xi_j=\partial\Xi\bigcap R^d_j$ of the hyper-rectangle's boundary with the configuration space $R^d_j$ of the j-th equation in (\ref{2i}). Each kernel $\tilde K_i$ in \eqref{3c} depends only on potential $V_j$ and can be found via the corresponding single particle equation from (\ref{2i}).

Using condition (\ref{2j}), we can finally obtain a transparent boundary condition for the wave function itself at the i-th facet of the hyper-rectangle, at moment $t$
\begin{widetext}
\begin{multline}
\frac{\partial\psi}{\partial x_i}=\int\limits_{0}^{t}\int\limits_{\partial\Xi_j}\tilde K_i(t_j,\xi,x_{d(j-1)+1}\ldots x_{i-1},a_i,x_{i+1}\ldots x_{dj}, \sigma_{d(j-1)+1}\ldots \sigma_{i-1},\sigma_{i+1}\ldots \sigma_{dj})\times\\
\left.\Phi(t_1,\ldots,t_{j-1},\xi,t_{j+1}\ldots t_M,x_1\ldots x_{d(j-1)},\sigma_{d(j-1)+1}\ldots \sigma_{i-1},a_i,\sigma_{i+1}\ldots \sigma_{dj},x_{dj+1},\ldots x_N)d\Sigma_j\,d\xi\right|_{\genfrac{}{}{0pt}{2}{t_m=t}{m=1,...,M}},
\label{3d}
\end{multline}
\end{widetext}
which expresses the normal derivative of the wave function at the i-th hyperfacet through the auxiliary function $\Phi$. The latter can be found by solving the remaining $M-1$ equations of (\ref{2i}), using equality (\ref{2j}) as the initial condition. The solution of the latter problem also requires a TBC that can be derived from the same equations (\ref{2i}). So, we have reduced the spatial non-locality of the TBC from $N$ dimensions to $d$ dimensions by partially hiding it inside the auxiliary function $\Phi$.

\subsection{External potential with a compact support}
We, so far, have not made any specific assumptions about the potentials $V_j$. Let us now assume that they have a compact support. This means that $V_j=0$, when $|x_i|\ge a_i$ for any $i$. (It is supposed that $d(j-1)\le i\le dj$.) In this case, outside the computational domain or at its boundary, each of the equations (\ref{2i}) is reduced to the $d$-dimensional free-space Schr\"odinger equation \eqref{11a}. By repeating the procedure of the previous section for each equation (\ref{2i}) when $|x_i|\ge a_i$, we can split them further into $d$ free-space 1D Schr\"odinger equations for new auxiliary functions $\Psi_j$. 

As it is known, the propagator $\Gamma$ of the 1D Schr\"odinger equation is
\begin{equation}
\Gamma(\mu,\nu)=\sqrt{\frac{ik}{2\pi \mu}}\exp\left(ik\frac{\nu^2}{2\mu}\right),
\label{31a}
\end{equation} 
where $\nu=x-x'$ and $\mu=t-t'$. Now the functions $\Psi_j$ can be obtained by convoluting $\Phi$ with $\Gamma$ $d-1$ times over the spatial variables of the respective equation from (\ref{2i}). All $\Psi_j$ functions will depend on $M+d-1$ temporal and $N$ spatial variables
\begin{widetext}
\begin{multline}
\Psi_j(\beta_{d(j-1)+1}\ldots\beta_{dj}, t_1\ldots t_{j-1}, t_{j+1}\ldots t_M, x_1\ldots x_N)=\\
\int\limits_{-\infty}^{+\infty}\ldots\int\limits_{-\infty}^{+\infty}\Phi(t_1\ldots t_{j-1}, \beta_j, t_{j+1}\ldots t_M,x_1\ldots x_{d(j-1)},\zeta_{d(j-1)+1}\ldots\zeta_{dj},x_{dj+1}\ldots x_N)\times\\
\prod\limits_{i=d(j-1)+1}^{dj-1}\Gamma(\beta_i-\beta_j,x_i-\zeta_i)d\zeta_i,
\label{31b}
\end{multline}
\end{widetext}
Each thus defined function $\Psi_j$ satisfies all of the equations (\ref{2i}) except the j-th one, while also satisfying $d$ 1D free-space Schr\"odinger equations \eqref{11a}, with respect to the temporal variables $\beta_{d(j-1)+1}\ldots\beta_{dj}$ and corresponding spatial variables $x_{d(j-1)+1}\ldots x_{dj}$. Function $\Psi_j$ is connected to $\psi$ by an equality similar to (\ref{2j}). 

Using these 1D Schr\"odinger equations and the Baskakov--Popov 1D transparent boundary condition (see \cite{baskakov1991implementation})
\begin{equation}
\frac{\partial \psi}{\partial x}=\mp\frac{1}{\sqrt{\pi i}}\frac{\partial}{\partial t}\int\limits_{0}^{t}\frac{\psi(\zeta,\pm a)}{\sqrt{t-\zeta}}\,d\zeta
\label{31c}
\end{equation}
it is possible to write down a TBC for the wave functions $\Psi_j$ in the following form
\begin{widetext}
\begin{equation}
\frac{\partial \Psi_j}{\partial x_i}=\int\limits_{0}^{t}\frac{\Psi_j(\beta_{d(j-1)+1}\ldots\beta_{i-1}, \zeta, \beta_{i+1}\ldots\beta_{dj}, t_1\ldots t_{j-1}, t_{j+1}\ldots t_M, x_1\ldots x_N)}{\sqrt{\beta_i-\zeta}}d\zeta.
\label{31c1}
\end{equation}
\end{widetext}
The final TBC for $\psi$ on the i-th hyperfacet will be as follows
\begin{widetext}
\begin{multline}
\frac{\partial\psi}{\partial x_i}=\left.\int\limits_{0}^{t}\frac{\Psi_j(\beta_{d(j-1)+1}\ldots\beta_{i-1}, \zeta, \beta_{i+1}\ldots\beta_{dj}, t_1\ldots t_{j-1}, t_{j+1}\ldots t_M, x_1\ldots x_N)}{\sqrt{\beta_i-\zeta}}d\zeta\right|_{\genfrac{}{}{0pt}{2}{\beta_{d(j-1)+s}=t}{s=1,...,d},\;\genfrac{}{}{0pt}{2}{\genfrac{}{}{0pt}{2}{t_m=t,}{m=1,...,M}}{m\ne j}},
\label{31d}
\end{multline}
\end{widetext}
Functions $\Psi_j$ appering in (\ref{31d}) at the boundary can be found by solving Schr\"odinger equations in the space with $N-1$ dimensions with appropriate TBCs. The latter boundary conditions will be similar to those derived above and will involve functions $\Psi_j$ with the number of spatial and temporal variables reduced by one. These functions, in turn, can be found by solving equations similar to (\ref{2i}) and 1D Schr\"odinger equations. This procedure needs to be repeated $N$ times, until the boundary conditions become one-dimensional. By using such a recurrent algorithm one can obtain a TBC for a system of any number of non-interacting particles in the N-dimensional space.

The boundary condition \eqref{31d} will eventually need to be realized as a part of some finite-difference scheme. In the case when the computational domain is shaped as an N-dimensional cube with all $a_l=a$, the FD scheme will have $N_x$ grid points along each spatial dimension and $N_\tau$ grid points along each time-like dimension. The resulting FD scheme can then be solved by various methods, such as the alternating direction implicit method (ADI) \cite{peaceman1955numerical}, or by directly inverting a large sparse matrix. In addition, the boundary values of the auxiliary functions $\Psi_j$ will need to be preserved between successive computational steps. The total number of stored values $NV$ will depend polynomially on the number of spatial and temporal steps in the finite-difference scheme as
\begin{equation}
NV=\sum_{s=0}^{N-1}D_s N_x^s N_\tau^{N-s},
\label{31e}
\end{equation}
where $D_s$ are coefficients equal to the total number of s-dimensional elements within the border $\partial\Xi$ of the computational domain. They can be expressed as
\begin{equation}
D_s=2^{N-s}d!/s!/(N-s)!
\label{31f}
\end{equation}
leading to the following final expression for $NV$
\begin{equation}
NV=(N_x+2N_\tau)^N-N_x^N.
\label{31g}
\end{equation}
The total complexity of any finite-difference scheme used to solve Schr\"odinger equation \eqref{2a} will also have a polynomial form similar to \eqref{31e} but with a more complex coefficients depending on the FD scheme used. 

As the general boundary condition formulated in this section relies on the Baskakov--Popov 1D TBC for its implementation, this 1D TBC \eqref{31c} will either need to be discretized as was done in \cite{feshchenko2011exact,feshchenko2013exact} or, instead, a new fully discrete TBC should be derived to be used in the FD scheme.

\section{Finite-difference schemes for the 3D and 4D cases}
\subsection{Fully discrete unconditionally stable 1D TBC}
Let us derive an exact fully discrete TBC (a brief description of this derivation can also be found in \cite{feshchenko2014parabolic}) for the Crank--Nicolson discretization \cite{PCP_Crank_1947} of the 1D Schr\"odinger equation in free space
\begin{equation}
i\frac{\partial \psi}{\partial t}+\frac{\partial^2\psi}{\partial x^2}=0.
\label{11a}
\end{equation}
The computational domain is defined as $|x|\leqslant a$, where $a$ is a positive real number. Equation (\ref{11a}) holds at least outside this domain and at its boundaries at $x=\pm a$. The initial condition for equation (\ref{11a}) is $\psi(x,t=0)=\psi_0(x)$, where $\psi_0$ is assumed to be equal to $\exp(iqx)$ if $|x|\geqslant a$ and $q$ is a parameter. The uniform Crank--Nicolson discretization of the equation (\ref{11a}) is as follows
\begin{multline}
-\psi^{n+1}_{p+1}+B\psi^{n+1}_{p}-\psi^{n+1}_{p-1}=\\
\psi^{n}_{p+1}-\tilde B\Psi^{n}_{p}+\psi^{n}_{p-1},
\label{11b}
\end{multline}
where
$$
B=2-2ih^2/\tau,\quad \tilde B=2+2ih^2/\tau,
$$
$$
0\leqslant p\leqslant N,\quad 0\leqslant n\leqslant N_\tau,
$$
$N+1$ is the total number of spatial grid nodes and $N_\tau+1$ is the total number of the temporal grid nodes. Parameters $h=2a/N$ and $\tau$ are the spatial and temporal steps, respectively. The boundary condition will be imposed at $x$ point corresponding to either $p=1$ or $p=N$.

Let us apply the one-sided \textit{Z}-transform
\begin{equation}
\Omega(p,w)=\sum\limits_{n=0}^{\infty}\frac{\psi^n_p}{w^n}
\label{11c}
\end{equation}
to the equation (\ref{11b}). The series in \eqref{11c} is considered convergent when $|z|>L$, where $L$ is a positive number. After some transformations and taking into account the initial condition for $\psi$ one can obtain the following equation for $\Omega$
\begin{multline}
\Omega(p+1,w)-2b\Omega(p,w)+\Omega(p-1,w)=\\
\frac{w}{1+w}(\psi^0_{p+1}-B\psi^0_{p}+\psi^0_{p-1}),
\label{11d}
\end{multline}
where
\begin{align}
b=&1+\frac{\alpha}{2}\frac{1-w}{1+w},\nonumber\\
\alpha=&\frac{2ih^2}{\tau},\quad \psi^0_p=e^{iqhp}.\label{11e}
\end{align}
The general solution of the equation (\ref{11d}) is
\begin{multline}
\Omega=C_1(w)\lambda_1^p+C_2(w)\lambda_2^p+\\
e^{iqhp}\frac{e^{iqh}-B+e^{-iqh}}{e^{iqh}-2b+e^{-iqh}}\frac{w}{1+w},
\label{11f}
\end{multline}
where $\lambda_{1,2}$ are the characteristic numbers of (\ref{11d}) that can be expressed as
\begin{equation}
\lambda_{1,2}=b\mp\sqrt{b^2-1}.
\label{11g}
\end{equation} 
The transparent (outgoing) boundary condition corresponds to $C_2=0$ or $C_1=0$ in (\ref{11f}) for the upper or lower boundary of the computational domain, respectively. This choice follows from the fact that $|b|>1$ for $w\to\infty$ and therefore $|\lambda_1|<1$ and $|\lambda_2|>1$ (note that $|\lambda_1\lambda_2|=1$). Now in order to obtain a discrete TBC, which is analogous to the discretized Baskakov--Popov condition, one needs to apply the inverse \textit{Z}-transform
\begin{equation}
\psi^n_p=\frac{1}{2\pi i}\oint\limits_{|w|=1}\Omega(p,w)w^{n-1}\,dw
\label{11h}
\end{equation}
to the difference $\Omega(N+1,w)-\Omega(N-1,w)$ (upper boundary) or $\Omega(2,w)-\Omega(0,w)$ (lower boundary). For instance, in the first case one obtains
\begin{multline}
\psi_{N+1}^{n+1}-\psi_{N-1}^{n+1}=\\
-2\sum_{l=0}^{n+1}\beta^{n+1-l}\psi_N^l+2ie^{iqhN}\sin(qh)g^{n+1}+\\
2e^{iqhN}\sum_{l=0}^{n+1}\beta^{n+1-l}g^{l},
\label{11i}
\end{multline}
where 
\begin{align}
\beta^n&=\frac{1}{2\pi i}\oint\limits_{|w|=1}\sqrt{b^2-1}w^{n-1}\,dw,\label{11j}\\
g^n&=\left(\frac{\alpha+2-2\cos qh}{\alpha-2+2\cos qh}\right)^n.\label{11j1}
\end{align}
For the lower boundary, the similar condition is (note the different sign)
\begin{multline}
\psi_{2}^{n+1}-\psi_{0}^{n+1}=\\
2\sum_{l=0}^{n+1}\beta^{n+1-l}\psi_1^l+2ie^{iqhN}\sin(qh)g^{n+1}-\\
2e^{iqhN}\sum_{l=0}^{n+1}\beta^{n+1-l}g^{l}.
\label{11k}
\end{multline}

Coefficients $\beta^n$ can be calculated using the following algorithm. First, it can be shown from (\ref{11j}) that
\begin{equation}
\beta^0=\sqrt{\frac{\alpha^2}{4}-\alpha}.
\label{11k1}
\end{equation}

Second, let us introduce function $f$ defined as
\begin{multline}
f(w)=\sqrt{b^2-1}=\\
i\sqrt{\alpha}\sqrt{\frac{1-1/w}{1+1/w}}\sqrt{1-\frac{\alpha}{4}\frac{1-1/w}{1+1/w}},
\label{11l}
\end{multline}
the logarithm of which is
\begin{multline}
\ln f(w)=const-\ln(1+1/w)+\frac{1}{2}\ln(1-1/w)+\\
\frac{1}{2}\ln\left(1+\frac{1+\alpha/4}{1-\alpha/4}\frac{1}{w}\right)=const+\\
\sum_{k=1}^{\infty}\left[-\frac{1}{2}-(-1)^{k-1}+\frac{1}{2}(-1)^{k-1}\left(\frac{1+\alpha/4}{1-\alpha/4}\right)^k\right]\frac{1}{kw^k}.
\label{11m}
\end{multline}
By differentiating (\ref{11m}) with respect to $w$, one obtains that
\begin{multline}
-w\frac{f'}{f}=\\
\sum_{k=1}^{\infty}\frac{1}{w^k}\left[-\frac{1}{2}-(-1)^{k-1}+\frac{1}{2}(-1)^{k-1}\left(\frac{1+\alpha/4}{1-\alpha/4}\right)^k\right].
\label{11n}
\end{multline}
Now, multiplying (\ref{11n}) by $f$, applying the inverse transform (\ref{11h}) to it and taking into account that $-wf'$ corresponds to $n\beta^n$ one can obtain a recurrent relation for the coefficients $\beta^n$ for any $n>0$
\begin{equation}
\beta^n=\frac{1}{n}\sum_{k=0}^{n-1}\varphi^{n-k}\beta^k,
\label{11o}
\end{equation}
where
\begin{multline}
\varphi^k=\frac{1}{2\pi i}\oint\limits_{|w|=1}w^{k-1}\times\\
\sum_{l=1}^{\infty}\frac{1}{w^l}\left[-\frac{1}{2}-(-1)^{l-1}+\frac{1}{2}(-1)^{l-1}\left(\frac{1+\alpha/4}{1-\alpha/4}\right)^l\right]\,dw=\\
-\frac{1}{2}-(-1)^{k-1}+\frac{1}{2}(-1)^{k-1}\left(\frac{1+\alpha/4}{1-\alpha/4}\right)^k.
\label{11p}
\end{multline}
From (\ref{11p}) it follows that
\begin{equation}
\begin{array}{rl}
\displaystyle\varphi^{2m+1}=&\displaystyle-\frac{3}{2}+\frac{1}{2}\left(\frac{1+\alpha/4}{1-\alpha/4}\right)^{2m+1},\\
\displaystyle\varphi^{2m}=&\displaystyle\frac{1}{2}\left[1-\left(\frac{1+\alpha/4}{1-\alpha/4}\right)^{2m}\right],
\end{array}
\label{11q}
\end{equation}
where $m=0,1,2,...$. The boundary conditions (\ref{11i}) and (\ref{11k}) can be rewritten as to separate the term proportional to $\psi_{N}^{n+1}$ from the sum as
\begin{align}
&\psi_{N+1}^{n+1}+2\beta^0\psi_{N}^{n+1}-\psi_{N-1}^{n+1}\nonumber\\
&=-2\sum_{l=0}^{n}\beta^{n+1-l}\psi_N^l+2ie^{iqhN}\sin(qh)g^{n+1}\nonumber\\
&+2e^{iqhN}\sum_{l=0}^{n+1}\beta^{n+1-l}g^{l},\label{11r1}\\
&\psi_{2}^{n+1}-2\beta^0\psi_{1}^{n+1}-\psi_{0}^{n+1}\nonumber\\
&=2\sum_{l=0}^{n}\beta^{n+1-l}\psi_1^l+2ie^{iqhN}\sin(qh)g^{n+1}\nonumber\\
&-2e^{iqhN}\sum_{l=0}^{n+1}\beta^{n+1-l}g^{l}.\label{11r2}
\end{align}
Discrete TBCs (\ref{11r1})--(\ref{11r2}) are similar to the fully discrete and semi-discrete boundary conditions derived in \cite{CommMathSci_Arnold_2003,yevick2001comparison}, respectively. The main differences are that (i) we use the second order approximation to the derivative by transversal coordinate $x$ and (ii) we consider a more general case when $\psi(x,0)=\exp(iqx)\ne0$. We also use a recurrent algorithm to calculate the convolution coefficients, which is more efficient than using explicit formulas.

\begin{figure}[!t]
\subfloat{%
\includegraphics[scale=0.25]{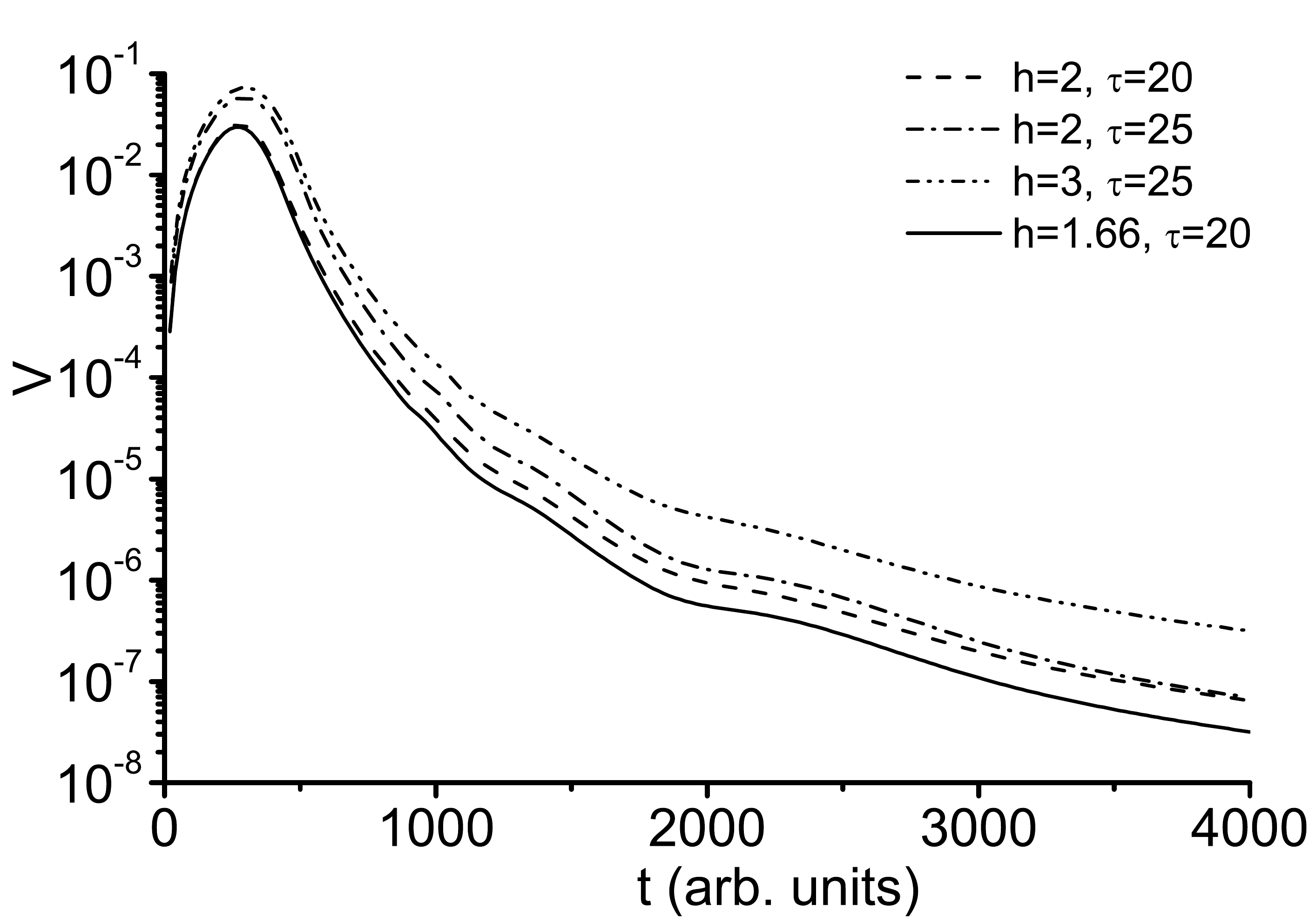}}

\subfloat{%
\includegraphics[scale=0.25]{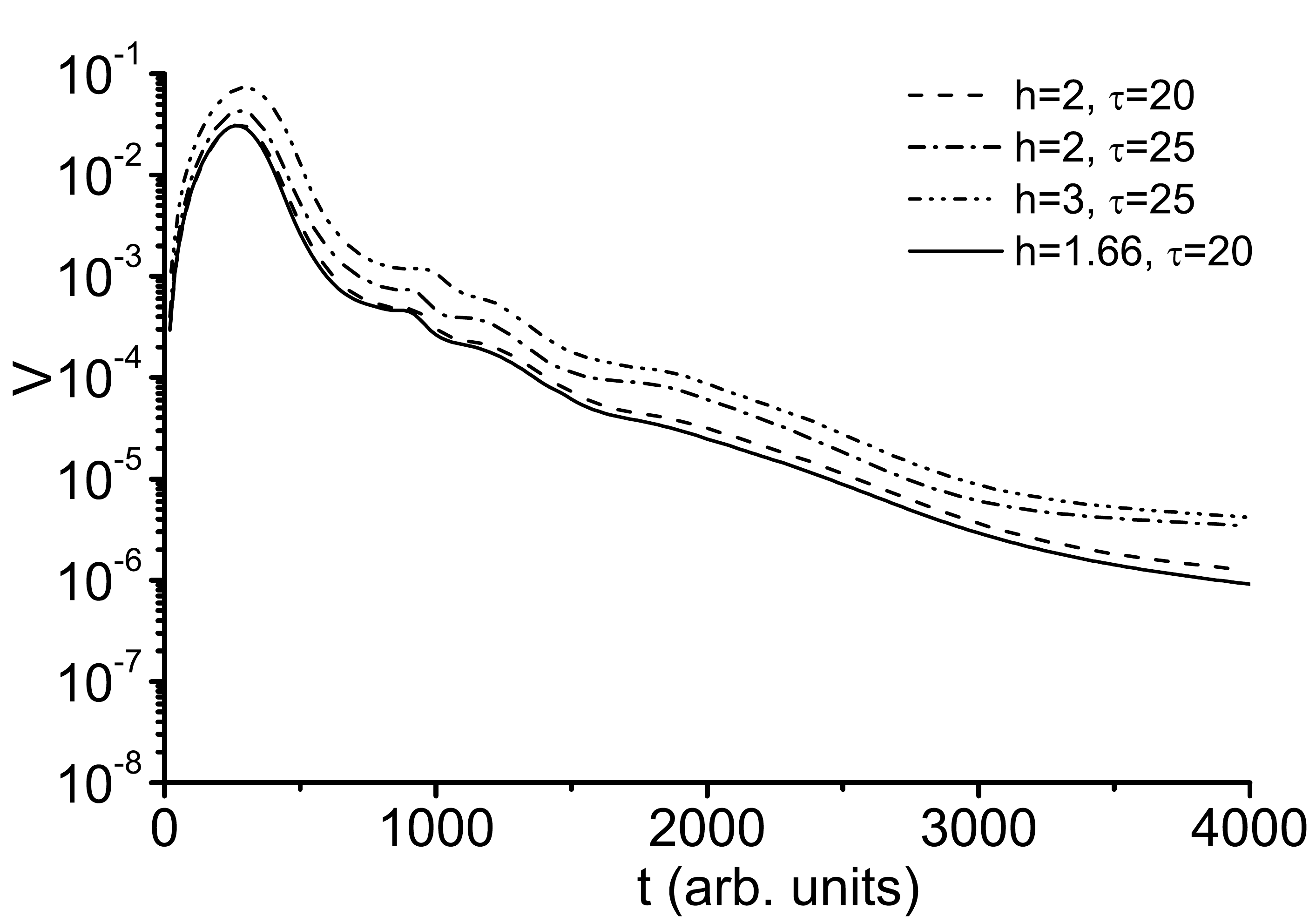}}
\caption{\small Propagation of a sum of Gaussian wave packets in free space. The curves show the difference $V$ (see (\ref{5c})) between the numerical and exact solutions for a number of values of parameters $\tau$ and $h$. The apparent maximum is caused by the discretization errors. The top figure corresponds the FD scheme with the fully discrete TBC, while the bottom one -- to the same FD scheme with the discretized Baskakov--Popov TBC.}
\label{f1}
\end{figure}

Coefficients $\beta^n$ have quite a remarkable behavior when $n\to\infty$. They do not go to zero as in the case of the discretization of the Baskakov--Popov condition but oscillate between $b^{2m}=\alpha$ and $b^{2m+1}=-\alpha$ ($m\to\infty$). This behavior can be explained by the existence of a pole at $w=-1$ in the expression $\sqrt{b^2-1}$ being, according to (\ref{11j}), a \textit{Z}-transform of $\beta^n$.

\subsection{3D TBC}
Let us now consider the case when $M=1$ and $d=N=3$. Then our multidimensional equation is a single particle time-dependent 3D Schr\"odinger equation 
\begin{equation}
i\frac{\partial \psi}{\partial t}+\frac{\partial^2\psi}{\partial x^2}+\frac{\partial^2\psi}{\partial y^2}+\frac{\partial^2\psi}{\partial z^2}-U(x,y,z)\psi=0,
\label{4a}
\end{equation}
where the potential $U$ is supposed to have a compact support and to belong to the $L^\infty$ space. At the initial moment $t=0$, the wave function $\psi=\psi_0(x,y,z)$ is supposed to vanish outside the computational domain much like the potential $U$. 

So, in this case the general multidimensional boundary condition derived above is applicable. The computational domain is supposed to be a cube with $a_1=a_2=a_3=a$. The standard Crank--Nicholson scheme can be used to solve \eqref{4a}. The computational grid is defined as follows. The computational steps in all spatial directions are set at $h=2a/N$ where $N+1$ is the number of grid points in each of spatial directions. The total number of points is $(N+1)^3$ or $(N+1)^3-12(N-1)-8$, if the eight vertices and twelve edges of the cube are excluded. The time step length is $\tau\ggg h$. The number of time steps is $N_{\tau}$. Now the finite difference approximation to (\ref{3a}) can be written as 
\begin{multline}
-\psi^{n+1}_{m+1,s,p}-\psi^{n+1}_{m,s+1,p}-\psi^{n+1}_{m,s,p+1}+\\
C^{n+1}_{m,s,p}\psi^{n+1}_{m,s,p}-\psi^{n+1}_{m-1,s,p}-\psi^{n+1}_{m,s-1,p}-\psi^{n+1}_{m,s,p-1}=\\
\psi^n_{m+1,s,p}+\psi^n_{m,s+1,p}+\psi^n_{m,s,p+1}-\\
\tilde C^n_{m,s,p}\psi^{n}_{m,s,p}+\psi^n_{m-1,s,p}+\psi^n_{m,s-1,p}+\psi^n_{m,s,p-1},
\label{4b}
\end{multline}
where
\begin{align}
&\psi^n_{m,s,p}=\psi(\tau n,hm,hs,hp),\label{4c1}\\
&U^n_{m,s,p}=U(\tau n,hm,hs,hp),\label{4c2}\\
&C^n_{m,s,p}=6+h^2U^n_{m,s,p}-2ih^2/\tau,\label{4c3}\\
&\tilde C^n_{m,s,p}=6+h^2U^n_{m,s,p}+2ih^2/\tau,\label{4c4}
\end{align}
and $0\leqslant n\leqslant N_{\tau}$, $1\leqslant m,s,p\leqslant N-1$ and $n$ is the current marching step. The boundaries of the computational domain are located at $m,s,p=N-1$ and $m,s,p=1$. Finite-difference scheme \eqref{4b} is solved by the inverting the large sparse matrix of the equations \eqref{4b} using an efficient iterative algorithm. The program code solving \eqref{4b} was implemented in the Matlab programming environment \cite{MatlabGithub}. 

Expressions \eqref{11r1} and \eqref{11r2} serve as the boundary conditions both for the main FD scheme \eqref{4b} as well as for the propagation of auxiliary function $\Psi$ at the facets and edges of the computational domain. Here it should be noted that the use the fully discrete boundary conditions \eqref{11r1} and \eqref{11r2} together with \eqref{4b} is not without controversy as the all considerations of the previous section are strictly speaking valid only for continuous functions. Therefore the conditions \eqref{11r1} and \eqref{11r2} and the finite-difference scheme \eqref{4b} might be somewhat incompatible with each other, especially in the corners. However, the numerical experiments reported below revealed that this problem is probably of a minor importance. Moreover, if the ADI were used there would likely be no incompatibility at all.

For the comparison the method of solution of \eqref{4b} based on the same iterative inversion of sparse matrices but with the standard discretization of the Baskakov--Popov boundary condition from \cite{feshchenko2013exact} is also used.

Let us also note that computational complexity $FC_3$ at each step of the scheme outlined above can be estimated by the following approximate relation
\begin{multline}
FC_3\approx P\times N^3+Q\times N^2\times n+\\
R\times N\times n^2+T\times n^3,
\label{4d}
\end{multline}
where $P$, $Q$, $R$ and $T$ are some coefficients. The first term in \eqref{4d} corresponds to the complexity of the sparse matrix inversion. The total number of values $NV$, which need to be stored between the computational steps, can be estimated by the formula \eqref{31e}, assuming $N=3$.

\section{4D TBC}
The solution of the two-particle 2D Schr\"odinger equation
\begin{multline}
i\frac{\partial \psi}{\partial t}+\frac{\partial^2\psi}{\partial x_1^2}+\frac{\partial^2\psi}{\partial y_1^2}+\frac{\partial^2\psi}{\partial x_2^2}+\\
\frac{\partial^2\psi}{\partial y_2^2}-(U_1(x_1,y_1)+U_2(x_2,y_2))\psi=0,
\label{4e}
\end{multline}
where $U_1$ and $U_2$ are some 2D potentials having compact support and to belong to the $L^\infty$ space, is of interest if one studies the evolution of entangled particle pairs in the 2D space \cite{dodd2004disentanglement,pivzorn2011time}. Equation \eqref{4e} corresponds to the general multidimensional case discussed in section \ref{secmult}) with the parameters set as $N=4$, $d=2$ and $M=2$. In addition, the coordinates in the equation \eqref{4e} have been normalized in order to consume the particle mass factors. 

The equation \eqref{4b}, being the Schr\"odinger equation in the 4D space, can be solved with the following finite difference scheme
\begin{multline}
-\psi^{n+1}_{m+1,s,p,\nu}-\psi^{n+1}_{m,s+1,p,\nu}-\psi^{n+1}_{m,s,p+1,\nu}+-\psi^{n+1}_{m,s,p,\nu+1}+\\
C^{n+1}_{m,s,p,\nu}\psi^{n+1}_{m,s,p,\nu}-\psi^{n+1}_{m-1,s,p,\nu}-\\
\psi^{n+1}_{m,s-1,p,\nu}-\psi^{n+1}_{m,s,p-1,\nu}-\psi^{n+1}_{m,s,p,\nu-1}=\\
\psi^n_{m+1,s,p,\nu}+\psi^n_{m,s+1,p,\nu}+\psi^n_{m,s,p+1,\nu}+\psi^n_{m,s,p,\nu+1}-\\
\tilde C^n_{m,s,p,\nu}\psi^{n}_{m,s,p,\nu}+\psi^n_{m-1,s,p,\nu}+\\
\psi^n_{m,s-1,p,\nu}+\psi^n_{m,s,p-1,\nu}+\psi^n_{m,s,p,\nu-1},
\label{4f}
\end{multline}
where
\begin{align}
&\psi^n_{m,s,p,\nu}=\psi(\tau n,hm,hs,hp,h\nu),\label{4g1}\\
&U^n_{m,s,p,\nu}=U(\tau n,hm,hs,hp,h\nu),\label{4g2}\\
&C^n_{m,s,p,\nu}=8+h^2U^n_{m,s,p,\nu}-2ih^2/\tau,\label{4g3}\\
&\tilde C^n_{m,s,p,\nu}=8+h^2U^n_{m,s,p,\nu}+2ih^2/\tau,\label{4g4}
\end{align}
and $0\leqslant n\leqslant N_{\tau}$, $1\leqslant m,s,p,\nu \leqslant N-1$ and $U=U_1+U_2$. 

The computational grid is defined similar to the 3D case. The computational steps in all spatial directions are set at $h=2a/N$ where $N+1$ is the number of grid points in each spatial direction. The total number of points is $(N+1)^4$ or $(N+1)^4-24(N-1)^2-32(N-1)-16$, if the 16 vertices, 32 edges and 24 facets of the hypercube are excluded. The time step size is $\tau\ggg h$. The number of time steps is $N_{\tau}$.

Let us also estimate the computational complexity $FC_4$ at each step of the FD scheme. By analogy with \eqref{4d} it will be
\begin{multline}
FC_4\approx P'\times N^4+Q'\times N^3\times n+R'\times N^2\times n^2+\\
S'\times N\times n^3+T'\times n^4,
\label{4i}
\end{multline}
where $P'$, $Q'$, $R'$, $S'$ and $T'$ are some coefficients. The total number of values $NV$, which must be stored between the computational steps, can be as well estimated with the formula \eqref{31e} assuming in this case that $N=4$.

\section{Numerical experiments}
\begin{table}
\begin{ruledtabular}
\begin{tabular}{r|c|c|c|l}
$\bm{l}$&$\bm{\cos\alpha}$&$\bm{\cos\beta}$&$\bm{\cos\gamma}$&$\bm{\xi}$\tabularnewline
\hline
1&$1/\sqrt{3}$&$1/\sqrt{3}$&$1/\sqrt{3}$&0.05\tabularnewline
\hline
2&$1/\sqrt{2}$&$1/\sqrt{2}$&0&0.1\tabularnewline
\hline
3&1&0&0&0.02
\end{tabular}
\end{ruledtabular}
\caption{The parameters of Gaussian wave packets used in (\ref{5a}).}
\label{t1}
\end{table}

\begin{figure*}
\includegraphics[scale=0.47]{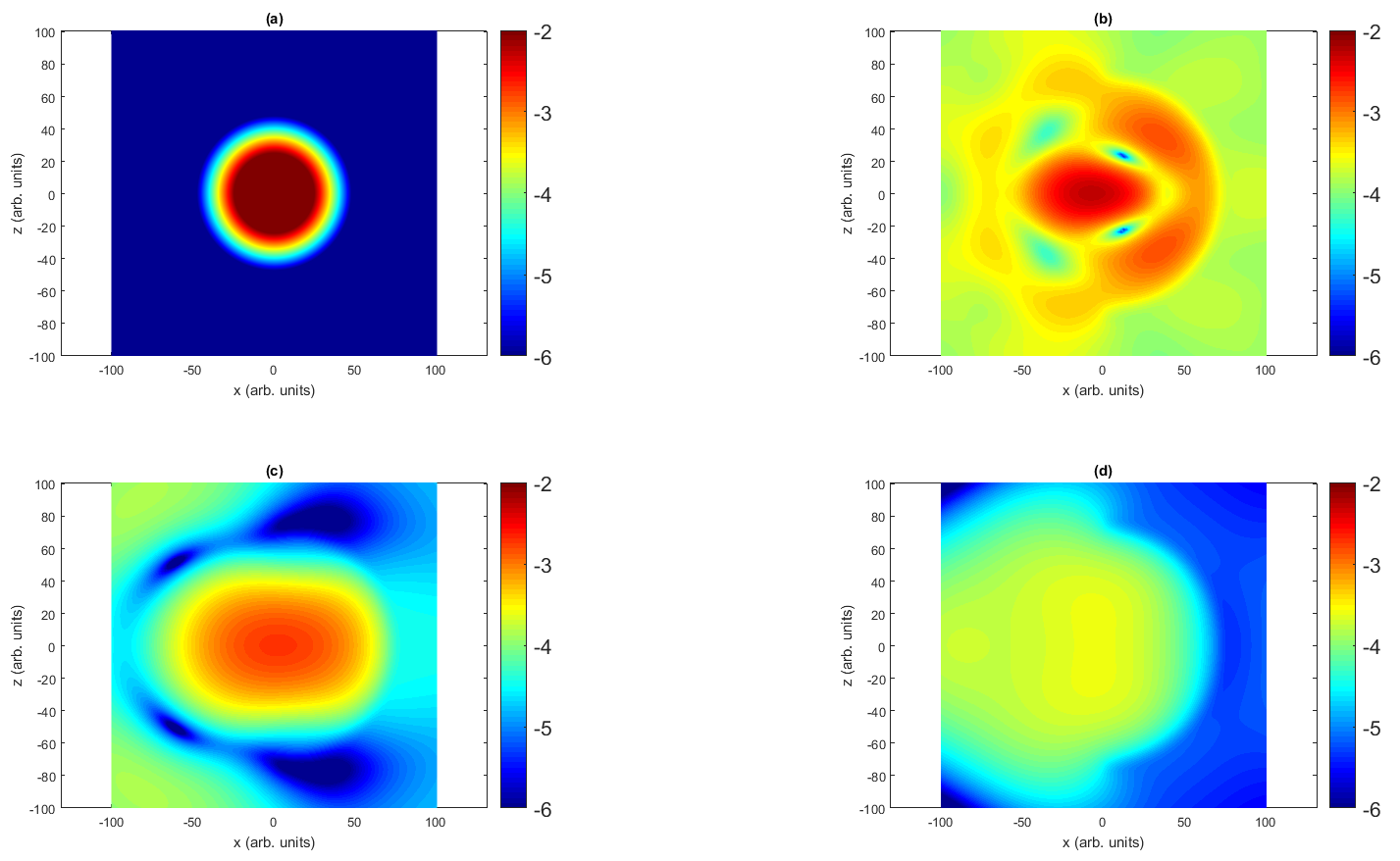}
\caption{\small (Color online) Evolution of wave function in a semi-spherical potential well with $r_0=70$. Distributions of $\log_{10}|\psi|^2$ in $(x,z)$ plane at $y=0$ is plotted at four different times:  (a) -- 320 arb. units, (b) -- 640 arb. units, (c) -- 1280 arb. units, (d) -- 1920 arb. units.}
\label{f2}
\end{figure*}
In this section, the validity of the derived boundary condition for beam propagation in 3D space is demonstrated by a number of numerical experiments from quantum mechanics, including evolution of a sum of Gaussian wave packets in free space and quantum tunneling through a 3D barrier. We do not conduct any 4D numerical experiments as this is beyond the capabilities of even the best available personal computers.

First, in order to quantitatively assess the accuracy of the proposed fully discrete and unconditionally stable TBC in comparison to the discretized Baskakov--Popov TBC we carried out a numerical experiment involving the propagation of a sum of 3D Gaussian wave packets having the following form
\begin{equation}
\psi_g(x,y,z,t)=\sum_{l=1}^{3}\psi^l_g(x,y,z,t).
\label{5a}
\end{equation}
Since any Gaussian packet is an exact solution of the Schr\"odinger equation, we can compare it with the results obtained using a finite-difference scheme with an appropriate TBC. The wave packets used here have the following form
\begin{widetext}
\begin{multline}
\psi^l_g(x,y,z,t)=\frac{w_l^3}{(w_l^2+4it)^{3/2}}\exp\left[i\xi_l((x-t\xi_l\cos\alpha_l)\cos\alpha_l+(y-t\xi_l\cos\beta_l)\cos\beta_l+(z-t\xi_l\cos\gamma_l)\cos\gamma_l)\right]\times\\
\exp\left[-((x-2t\xi_l\cos\alpha_l)^2+(y-2t\xi_l\cos\beta_l)^2+(z-2t\xi_l\cos\gamma_l)^2)/(w_l^2+4it)\right],
\label{5b}
\end{multline}
\end{widetext}
where $\cos\alpha_{x,y,z}$  are the direction cosines and $\xi_l$ are coefficients, whose values can be found in Table \ref{t1}. 

The waist radius $w$ of all packets is taken as 18. The size of the computational domain is $200\times200\times200$ with the grid step equal to 3, 2 or 1.66. The temporal step is 25 or 20, with the number of steps equal to 200. The initial condition is chosen as $u_0(x,y,z)=u_g(0,x,y,z)$. For each numerical experiment, relative discrepancy $V$ between the exact solution (\ref{5b}) and the finite-difference approximation is calculated and shown as a function of $t$. Function $V$ is defined as
\begin{multline}
V(\tau n)=\sum\limits_{m,s,p=0}^{N}|\psi^n_{m,s,p}-\psi_g(\tau n,hm,hs,hp)|^2/\\
\sum\limits_{m,s,p=0}^{N}|\psi_g(\tau n,hm,hs,hp)|^2.
\label{5c}
\end{multline}
The results for a number of $h$, $\tau$ are plotted in Fig. \ref{f1} calculated by two methods (upper and lower sub-figures, respectively). The peak near $t=300$ observed on all curves is caused by a discretization error of the finite-difference scheme and does not depend significantly on the type of transparent boundary condition used. The secondary smaller peaks  near 1000--1500, and especially prominent for the discretized Baskakov--Popov TBC in the bottom Fig. \ref{f1}, are caused by spurious reflections from the artificial boundary, due to discretization of the TBC. Beyond these maximums all curves monotonically go to zero without any obvious instabilities. It is clear from Fig. \ref{f1} that the sparse matrix based FD scheme with the fully discrete 1D TBC, asymptotically (when $t\to\infty$) provides much better -- by a several orders of magnitude -- precision compared with the same FD scheme with the discretized Baskakov--Popov TBC.

As the second example we will consider reflection of a particle off the 3D semi-spherical potential defined as a simple shell-like barrier with $U=0$ for $r<r_0$ or $r>r_1$ or $x<0$ and $U=0.01$ if $r_0<r<r_1$ and $x\gg0$. In this simulation the size of computational domain was again $200\times 200\times 200$ units whereas other parameters were: $h=1.66$, $\tau=20$ and $N_\tau=200$. The initial $\psi$ distribution at $t=0$ was chosen as a Gaussian packet (\ref{5b}) with the waist radius of $w=14$ and all $\xi_l=0$. This initial $\psi$ field was propagated by the sparse matrix-based FD scheme with a fully discrete 1D TBC.

The results of the simulation are shown in Fig.\ref{f2} as four successive quantum state densities $|\psi|^2$ in the $(x,z)$ plane. In this numerical experiment the semi-shell acts like a concave mirror rejecting and focusing a fraction of the $\psi$ field towards negative $x$ direction.

\section{Conclusion}
In this article, we have formulated a general exact multi-dimensional transparent boundary condition for the multi-particle Shr\"odinger equations in a hyperrectangular computational domain with a separable potential of quite a general form. We demonstrate that it can be reduced to transparent boundary conditions of a lower dimension for an auxiliary function with multiple time-like arguments. We have shown that in case of the potential having a compact support it can be further reduced to the Baskakov--Popov 1D boundary condition for this auxiliary function. 

The practical implementation of the obtained boundary condition is possible with different finite-difference schemes and requires either discretization of the Baskakov--Popov 1D TBC or the use of an equivalent fully discrete 1D boundary condition. We have obtained such an exact fully discrete boundary condition for the Crank--Nicholson FD scheme using the method of \textit{Z}-transforms. Since the FD implementation of the obtained multi-dimensional TBC requires memorizing the boundary values of the auxiliary function between computational steps, we estimated the number values that need to be stored and the computational complexity of the solution.

We considered the finite-difference implementation of the derived boundary condition in the case of a one-particle system in three dimensions as well as of a two-particle system in 2D. In the 3D case we demonstrate that the iterative use of sparse matrix inversion, together with the fully discrete 1D TBC, significantly improves the precision of the FD method used. In a two-particle case, we have not attempted any numerical experiments as they would be prohibitively computationally expensive. 

To resume, the proposed multi- and one-dimensional boundary conditions are simple, robust, stable and can be useful in the field of computational quantum mechanics, when an exact solution of the multi-dimensional Schr\"odinger equation, such as multiparticle problems, is required.

\section{Acknowledgments}
The authors are indebted to A.V. Vinogradov for the fruitful discussions on quantum mechanical problems. 

\bibliographystyle{apsrev4-2} 
\bibliography{PhysREvE_multi}

\begin{thebibliography}{15}%
\makeatletter
\providecommand \@ifxundefined [1]{%
 \@ifx{#1\undefined}
}%
\providecommand \@ifnum [1]{%
 \ifnum #1\expandafter \@firstoftwo
 \else \expandafter \@secondoftwo
 \fi
}%
\providecommand \@ifx [1]{%
 \ifx #1\expandafter \@firstoftwo
 \else \expandafter \@secondoftwo
 \fi
}%
\providecommand \natexlab [1]{#1}%
\providecommand \enquote  [1]{``#1''}%
\providecommand \bibnamefont  [1]{#1}%
\providecommand \bibfnamefont [1]{#1}%
\providecommand \citenamefont [1]{#1}%
\providecommand \href@noop [0]{\@secondoftwo}%
\providecommand \href [0]{\begingroup \@sanitize@url \@href}%
\providecommand \@href[1]{\@@startlink{#1}\@@href}%
\providecommand \@@href[1]{\endgroup#1\@@endlink}%
\providecommand \@sanitize@url [0]{\catcode `\\12\catcode `\$12\catcode
  `\&12\catcode `\#12\catcode `\^12\catcode `\_12\catcode `\%12\relax}%
\providecommand \@@startlink[1]{}%
\providecommand \@@endlink[0]{}%
\providecommand \url  [0]{\begingroup\@sanitize@url \@url }%
\providecommand \@url [1]{\endgroup\@href {#1}{\urlprefix }}%
\providecommand \urlprefix  [0]{URL }%
\providecommand \Eprint [0]{\href }%
\providecommand \doibase [0]{https://doi.org/}%
\providecommand \selectlanguage [0]{\@gobble}%
\providecommand \bibinfo  [0]{\@secondoftwo}%
\providecommand \bibfield  [0]{\@secondoftwo}%
\providecommand \translation [1]{[#1]}%
\providecommand \BibitemOpen [0]{}%
\providecommand \bibitemStop [0]{}%
\providecommand \bibitemNoStop [0]{.\EOS\space}%
\providecommand \EOS [0]{\spacefactor3000\relax}%
\providecommand \BibitemShut  [1]{\csname bibitem#1\endcsname}%
\let\auto@bib@innerbib\@empty
\bibitem [{\citenamefont {Antoine}\ \emph {et~al.}(2008)\citenamefont
  {Antoine}, \citenamefont {Arnold}, \citenamefont {Besse}, \citenamefont
  {Ehrhardt},\ and\ \citenamefont {Sch\"adle}}]{ComCompPhys_Antoine_2008}%
  \BibitemOpen
  \bibfield  {author} {\bibinfo {author} {\bibfnamefont {X.}~\bibnamefont
  {Antoine}}, \bibinfo {author} {\bibfnamefont {A.}~\bibnamefont {Arnold}},
  \bibinfo {author} {\bibfnamefont {C.}~\bibnamefont {Besse}}, \bibinfo
  {author} {\bibfnamefont {M.}~\bibnamefont {Ehrhardt}},\ and\ \bibinfo
  {author} {\bibfnamefont {A.}~\bibnamefont {Sch\"adle}},\ }\href@noop {}
  {\bibfield  {journal} {\bibinfo  {journal} {Commun. Comput. Phys.}\ }\textbf
  {\bibinfo {volume} {4}},\ \bibinfo {pages} {729} (\bibinfo {year}
  {2008})}\BibitemShut {NoStop}%
\bibitem [{\citenamefont {Popov}(1996)}]{popov1996accurate}%
  \BibitemOpen
  \bibfield  {author} {\bibinfo {author} {\bibfnamefont {A.~V.}\ \bibnamefont
  {Popov}},\ }\href {https://doi.org/10.1029/96RS02538} {\bibfield  {journal}
  {\bibinfo  {journal} {Radio Science}\ }\textbf {\bibinfo {volume} {31}},\
  \bibinfo {pages} {1781} (\bibinfo {year} {1996})}\BibitemShut {NoStop}%
\bibitem [{\citenamefont {Heinen}\ and\ \citenamefont
  {Kull}(2009)}]{PhysRev_Heinen_2009}%
  \BibitemOpen
  \bibfield  {author} {\bibinfo {author} {\bibfnamefont {M.}~\bibnamefont
  {Heinen}}\ and\ \bibinfo {author} {\bibfnamefont {H.-J.}\ \bibnamefont
  {Kull}},\ }\href {https://doi.org/10.1103/PhysRevE.79.056709} {\bibfield
  {journal} {\bibinfo  {journal} {Phys. Rev. E}\ }\textbf {\bibinfo {volume}
  {79}},\ \bibinfo {pages} {056709} (\bibinfo {year} {2009})}\BibitemShut
  {NoStop}%
\bibitem [{\citenamefont {Feshchenko}\ and\ \citenamefont
  {Popov}(2011)}]{feshchenko2011exact}%
  \BibitemOpen
  \bibfield  {author} {\bibinfo {author} {\bibfnamefont {R.}~\bibnamefont
  {Feshchenko}}\ and\ \bibinfo {author} {\bibfnamefont {A.}~\bibnamefont
  {Popov}},\ }\href {https://doi.org/10.1364/JOSAA.28.000373} {\bibfield
  {journal} {\bibinfo  {journal} {JOSA A}\ }\textbf {\bibinfo {volume} {28}},\
  \bibinfo {pages} {373} (\bibinfo {year} {2011})}\BibitemShut {NoStop}%
\bibitem [{\citenamefont {Feshchenko}\ and\ \citenamefont
  {Popov}(2013)}]{feshchenko2013exact}%
  \BibitemOpen
  \bibfield  {author} {\bibinfo {author} {\bibfnamefont {R.}~\bibnamefont
  {Feshchenko}}\ and\ \bibinfo {author} {\bibfnamefont {A.}~\bibnamefont
  {Popov}},\ }\href {https://doi.org/10.1103/PhysRevE.88.053308} {\bibfield
  {journal} {\bibinfo  {journal} {Physical Review E}\ }\textbf {\bibinfo
  {volume} {88}},\ \bibinfo {pages} {053308} (\bibinfo {year}
  {2013})}\BibitemShut {NoStop}%
\bibitem [{\citenamefont {Brandt}\ and\ \citenamefont
  {Dahmen}(1991)}]{brandt1991free}%
  \BibitemOpen
  \bibfield  {author} {\bibinfo {author} {\bibfnamefont {S.}~\bibnamefont
  {Brandt}}\ and\ \bibinfo {author} {\bibfnamefont {H.~D.}\ \bibnamefont
  {Dahmen}},\ }in\ \href@noop {} {\emph {\bibinfo {booktitle} {Quantum
  Mechanics on the Macintosh{\textregistered}}}}\ (\bibinfo  {publisher}
  {Springer},\ \bibinfo {year} {1991})\ pp.\ \bibinfo {pages}
  {109--130}\BibitemShut {NoStop}%
\bibitem [{\citenamefont {Baskakov}\ and\ \citenamefont
  {Popov}(1991)}]{baskakov1991implementation}%
  \BibitemOpen
  \bibfield  {author} {\bibinfo {author} {\bibfnamefont {V.~A.}\ \bibnamefont
  {Baskakov}}\ and\ \bibinfo {author} {\bibfnamefont {A.~V.}\ \bibnamefont
  {Popov}},\ }\href {https://doi.org/10.1016/0165-2125(91)90053-Q} {\bibfield
  {journal} {\bibinfo  {journal} {Wave motion}\ }\textbf {\bibinfo {volume}
  {14}},\ \bibinfo {pages} {123} (\bibinfo {year} {1991})}\BibitemShut
  {NoStop}%
\bibitem [{\citenamefont {Peaceman}\ and\ \citenamefont
  {Rachford}(1955)}]{peaceman1955numerical}%
  \BibitemOpen
  \bibfield  {author} {\bibinfo {author} {\bibfnamefont {D.~W.}\ \bibnamefont
  {Peaceman}}\ and\ \bibinfo {author} {\bibfnamefont {H.~H.}\ \bibnamefont
  {Rachford}, \bibfnamefont {Jr}},\ }\href {https://doi.org/10.1137/0103003}
  {\bibfield  {journal} {\bibinfo  {journal} {Journal of the Society for
  industrial and Applied Mathematics}\ }\textbf {\bibinfo {volume} {3}},\
  \bibinfo {pages} {28} (\bibinfo {year} {1955})}\BibitemShut {NoStop}%
\bibitem [{\citenamefont {Feshchenko}\ and\ \citenamefont
  {Popov}(2014)}]{feshchenko2014parabolic}%
  \BibitemOpen
  \bibfield  {author} {\bibinfo {author} {\bibfnamefont {R.}~\bibnamefont
  {Feshchenko}}\ and\ \bibinfo {author} {\bibfnamefont {A.}~\bibnamefont
  {Popov}},\ }in\ \href {https://doi.org/10.1007/978-3-319-00696-3_17} {\emph
  {\bibinfo {booktitle} {X-Ray Lasers 2012}}}\ (\bibinfo  {publisher}
  {Springer},\ \bibinfo {year} {2014})\ pp.\ \bibinfo {pages}
  {105--108}\BibitemShut {NoStop}%
\bibitem [{\citenamefont {Crank}\ and\ \citenamefont
  {Nicolson}(1947)}]{PCP_Crank_1947}%
  \BibitemOpen
  \bibfield  {author} {\bibinfo {author} {\bibfnamefont {J.}~\bibnamefont
  {Crank}}\ and\ \bibinfo {author} {\bibfnamefont {P.}~\bibnamefont
  {Nicolson}},\ }\href {https://doi.org/10.1017/S0305004100023197} {\bibfield
  {journal} {\bibinfo  {journal} {Proc. Cambridge Philos. Soc.}\ }\textbf
  {\bibinfo {volume} {43}},\ \bibinfo {pages} {50} (\bibinfo {year}
  {1947})}\BibitemShut {NoStop}%
\bibitem [{\citenamefont {Arnold}\ \emph {et~al.}(2003)\citenamefont {Arnold},
  \citenamefont {Ehrhardt},\ and\ \citenamefont
  {Sofronov}}]{CommMathSci_Arnold_2003}%
  \BibitemOpen
  \bibfield  {author} {\bibinfo {author} {\bibfnamefont {A.}~\bibnamefont
  {Arnold}}, \bibinfo {author} {\bibfnamefont {M.}~\bibnamefont {Ehrhardt}},\
  and\ \bibinfo {author} {\bibfnamefont {I.}~\bibnamefont {Sofronov}},\
  }\href@noop {} {\bibfield  {journal} {\bibinfo  {journal} {Comm. Math. Sci.}\
  }\textbf {\bibinfo {volume} {1}},\ \bibinfo {pages} {501} (\bibinfo {year}
  {2003})}\BibitemShut {NoStop}%
\bibitem [{\citenamefont {Yevick}\ \emph {et~al.}(2001)\citenamefont {Yevick},
  \citenamefont {Friese},\ and\ \citenamefont
  {Schmidt}}]{yevick2001comparison}%
  \BibitemOpen
  \bibfield  {author} {\bibinfo {author} {\bibfnamefont {D.}~\bibnamefont
  {Yevick}}, \bibinfo {author} {\bibfnamefont {T.}~\bibnamefont {Friese}},\
  and\ \bibinfo {author} {\bibfnamefont {F.}~\bibnamefont {Schmidt}},\ }\href
  {https://doi.org/10.1006/jcph.2001.6708} {\bibfield  {journal} {\bibinfo
  {journal} {Journal of Computational Physics}\ }\textbf {\bibinfo {volume}
  {168}},\ \bibinfo {pages} {433} (\bibinfo {year} {2001})}\BibitemShut
  {NoStop}%
\bibitem [{\citenamefont {Feshchenko}(2021)}]{MatlabGithub}%
  \BibitemOpen
  \bibfield  {author} {\bibinfo {author} {\bibfnamefont {R.}~\bibnamefont
  {Feshchenko}},\ }\href@noop {} {\bibinfo {title} {Matlab{TBC}}},\ \bibinfo
  {howpublished} {\url{https://github.com/Rusjava/MatlabTBC}} (\bibinfo {year}
  {2021})\BibitemShut {NoStop}%
\bibitem [{\citenamefont {Dodd}\ and\ \citenamefont
  {Halliwell}(2004)}]{dodd2004disentanglement}%
  \BibitemOpen
  \bibfield  {author} {\bibinfo {author} {\bibfnamefont {P.}~\bibnamefont
  {Dodd}}\ and\ \bibinfo {author} {\bibfnamefont {J.}~\bibnamefont
  {Halliwell}},\ }\href {https://doi.org/10.1103/PhysRevA.69.052105} {\bibfield
   {journal} {\bibinfo  {journal} {Physical Review A}\ }\textbf {\bibinfo
  {volume} {69}},\ \bibinfo {pages} {052105} (\bibinfo {year}
  {2004})}\BibitemShut {NoStop}%
\bibitem [{\citenamefont {Pi{\v{z}}orn}\ \emph {et~al.}(2011)\citenamefont
  {Pi{\v{z}}orn}, \citenamefont {Wang},\ and\ \citenamefont
  {Verstraete}}]{pivzorn2011time}%
  \BibitemOpen
  \bibfield  {author} {\bibinfo {author} {\bibfnamefont {I.}~\bibnamefont
  {Pi{\v{z}}orn}}, \bibinfo {author} {\bibfnamefont {L.}~\bibnamefont {Wang}},\
  and\ \bibinfo {author} {\bibfnamefont {F.}~\bibnamefont {Verstraete}},\
  }\href {https://doi.org/10.1103/PhysRevA.83.052321} {\bibfield  {journal}
  {\bibinfo  {journal} {Physical Review A}\ }\textbf {\bibinfo {volume} {83}},\
  \bibinfo {pages} {052321} (\bibinfo {year} {2011})}\BibitemShut {NoStop}%
\end{thebibliography}%

\end{document}